\documentclass[useAMS,usenatbib]{mn2e}
\usepackage{graphics}
\usepackage{graphicx}
\usepackage{aas_macros}
\usepackage{amsfonts}
\usepackage{amsmath}
\usepackage{color}

\newcommand{\beq}{\begin{equation}}
\newcommand{\enq}{\end{equation}}
\newcommand{\beqa}{\begin{eqnarray}}
\newcommand{\enqa}{\end{eqnarray}}
\newcommand{\beit}{\begin{itemize}}
\newcommand{\enit}{\end{itemize}}
\newcommand{\bem}{\begin{pmatrix}}
\newcommand{\enm}{\end{pmatrix}}

\newcommand{\vecr}{\mathbf{r}}

\newcommand{\lat}{\left\langle}
\newcommand{\rat}{\right\rangle}
\newcommand{\av}[1]{\lat #1 \rat}

\newcommand{\lb}{\left [}
\newcommand{\rb}{\right ]}
\newcommand{\lp}{\left (}
\newcommand{\rp}{\right )}

\renewcommand{\bem}{\begin{bmatrix}}
\renewcommand{\enm}{\end{bmatrix}}
\newcommand{\vecn}{\mathbf n}
\newcommand{\vecm}{\mathbf m}
\newcommand{\T}{\tilde M}
\renewcommand{\t}{p_W}
\newcommand{\tlambda}{p_W(\theta)}

\newcommand{\revision}[1]{#1}

\title[N-point correlators of the lognormal field]{What does the N-point function hierarchy of the cosmological matter density field really measure ?}
\author[Carron and Szapudi]{J. Carron$^{1}$ and I. Szapudi$^{2}$
\\$^{1}$ Department of Physics and Astronomy, University of Sussex, Brighton BN1 9QH, UK
\\$^{2}$ Institute for Astronomy, University of Hawaii, 96822 Honolulu,HI, USA. }
\begin{document}

\date{\today}

\pagerange{\pageref{firstpage}--\pageref{lastpage}} \pubyear{2016}

\maketitle

\label{firstpage}

\begin{abstract}
The cosmological dark matter field is not completely described by its hierarchy of $N$-point functions, a non-perturbative effect with the consequence that only part of the theory can be probed with the hierarchy. We give here an exact characterization of the joint information of the hierarchy within the lognormal field. The lognormal field is the archetypal example of a field where this effect occurs, and, at the same time, one of the few tractable and insightful available models to specify fully the statistical properties of the evolved matter density field beyond the perturbative regime.  Nonlinear growth in the Universe in that model is set letting the log-density field probability density functional evolve keeping its Gaussian shape, according to the diffusion equation in Euclidean space.  We show that the hierarchy probes a different evolution equation, the diffusion equation defined not in Euclidean space but on the compact torus, with uniformity as the long-term solution. The extraction of the hierarchy of correlators can be recast in the form of a nonlinear transformation applied to the field, 'wrapping', undergoing a sharp transition towards complete disorder in the deeply nonlinear regime, where all memory of the initial conditions is lost.
\end{abstract}

\begin{keywords}{large-scale structure of Universe, cosmology: theory, methods: statistical} 
\end{keywords}

\section{Overview}
According to the successful cosmological $\Lambda$ cold dark matter ($\Lambda$CDM) model, the evolution of the matter density field can be represented by the Vlasov-Poisson system in an expanding Universe, starting from Gaussian initial conditions. The study of this nonlinear system is a complex endeavour, typically addressed with perturbation theory (PT) in various forms or $N$-body numerical simulations nowadays, motivations being not only better understanding of structure formation, but also the fact that most cosmological probes of the low redshift Universe and of its accelerated expansion \citep{Albrecht:2006um,WeinbergEtal2013} must rely,  in order to fulfil their promises, on a thorough qualitative and quantitative understanding of the field's statistical properties.
 \newline
\indent
Complete specification of a field statistics requires a probability density 'functional', a self-consistent prescription assigning probabilities to joint occurrences at all points (or, for all practical purposes, at the different resolution elements, or 'pixels'). In the cosmological framework this functional is given formally by an intractable Liouville equation, describing conservation of probability along trajectories in phase space. Perturbative methods such as PT can trace successfully some simple properties of that functional close to the linear regime,  such as two-point statistics, while $N$-body numerical methods provide single field realizations only. Insights gained from greatly simplified models can thus often prove invaluable. The lognormal density field \citep{ColesJones1991} is one of these models. It is defined as the exponential of a Gaussian random field. The Gaussian log-density field is fully specified by requesting the resulting density two-point function to match that of the cosmological model.
\newline
\indent One wonders how a parameter-free, local transformation of a Gaussian field can possibly explain many qualitative properties of the evolving matter field. 
The most compelling argument at the time of writing is probably that this field exhibits hierarchical form \citep{Fry1984b,Peebles1980} in the weakly nonlinear regime of structure formation, and reduces naturally to a Gaussian field at early times, as required from the inflationary paradigm of modern cosmology. Typical hierarchical amplitudes  such as the time-honored one-point cumulants $S_p$  \citep{Bernardeau1994} do match well on these scales those expected in a $\Lambda$CDM Universe, showing consistency with PT. Early works have shown that the lognormal one-point probability distribution function (PDF) is thus a good match to that observed in simulations \citep[e.g.]{BernardeauKofman1995,KayoEtal2001}. More recently, the lognormal field was proved successful at modeling much more elaborated properties, such as the coupling of very large wavelength perturbations to small scales Fourier modes \citep{Takahashi:2014yqa,CarronSzapudi2015}. The ease with which it can be simulated makes it also a most practical and useful tool to simulate covariance matrices for cosmological data, 
and it is successful as a prior in bayesian matter field reconstruction from large scale structure data. \citep{HilbertEtal2011,KitauraEtal10,CarronEtal2014,AtaEtal2015}.
\newline
\indent
One of the most intriguing property of the lognormal field $\delta$ is that the full set of the in cosmology most traditional statistics the $N$-point functions \citep{Peebles1980,Fry1984b,BernardeauEtal2002,Szapudi2009}
\beq
\xi_N(x_1,\cdots,x_N) = \av{\delta(x_1)\cdots \delta(x_N)}_{\textrm{c}}
\enq
or their Fourier transform the polyspectra does not provide a complete statistical description of the field, as already pointed out in \citep{ColesJones1991}.  This is generic behavior for skewed nonlinear fields in which the series expansion of the moment generating function is only asymptotic \citep{Carron2011}, as is the case for the cosmological matter density field. We refer the reader to  \citep{ValageasII} for a very detailed study of the density field moment generating function; the main idea of how this can be is fairly simple: given the observed non-linear density $\delta$ one may write schematically the moment generating function $\av{e^{-t \delta}}$ going back to the initial, Gaussian, linear density as
\beq
\av{e^{-t \delta}} = \int \mathcal D \delta_L \:p[\delta_L] e^{- t \delta[\delta_L]}.
\enq
If nonlinear growth $\delta[\delta_L]$ more than compensates the Gaussian decay $-\delta_L^2$ in the exponent, then the above expression diverges for any $t <0$. This prevents the convergence of the series $\av{e^{-t \delta}} \sim \sum_n(-t)^n\av{\delta^n} / n!$ for any $t$ however small, and thus the recovery of $p(\delta)$ from the moments.
\newline
\indent For such fields, polynomials in the field do not form a complete basis set of functions and thus only part of the theory is probed by the hierarchy. It follows that there exists a variety of probability functionals that predict the same hierarchy of correlators, some of which given by \citep{CarronNeyrinck2012}, and information escapes the hierarchy. A deeper physical interpretation of this effect remains mysterious however, and it is not entirely clear what it is that the hierarchy does and does not characterize.
It is the purpose of this Letter to elucidate these questions in the case of the lognormal field. 
\newline
\indent
We show here that the hierarchy (in fact, a somewhat larger set, defined below, allowing in effect the use of Laurent-series in place of only ordinary power series in the field) does measure the probability functional not of the Gaussian log-density field, but that of a different random field, the 'wrapped' Gaussian field.  The time evolution of both functionals is given by the same equation, the diffusion equation, with identical initial conditions. The former diffuses in Euclidean space keeping its Gaussian shape, the latter however on the compact torus, reaching homogeneity in the long-term, the information contained in the hierarchy diffusing to zero quite literally. \revision{Eventually, this will allow us, as our main result in Eq.~\ref{decomp}, to characterize  the hierarchy of moments as capturing only a specific part of the log-density field. In order to do this, we will only need derive the hierarchy information matrix and comparing it to that of a field different to the log-density.} 
\newline
\indent
\textit{Gaussian probability functional and diffusion equation}
One way to characterize the Gaussian regime, thus, linear growth in the Universe, is as already advertised through a diffusion process. Let $p_G$ denote the Gaussian PDF for the fluctuation field $\delta$ at any number $d$ of different spatial positions. We use for simplicity as 'time' coordinate $t$ the cosmological growth factor $D^2$. Using the vector notation $\delta = (\delta(x_1),\cdots,\delta(x_d))$, $p_G$ can be explicitly written as
\beq
p_G(\delta,t) = \frac{ \exp\lp-\frac 1{2t}\delta^T \xi^{-1} \delta \rp}{(2\pi)^{d/2}\sqrt{\det {t \xi}}},
\enq
with covariance matrix $\xi_{ij} = \xi(x_i-x_j)$ the linear two-point function of the field (at $t = 1$). 
The time evolution of $p_G$ is given by the diffusion equation
\beq
\frac{\partial p_G}{\partial t} = \frac 12 \sum_{i,j} \xi_{ij}\frac{ \partial^2 p_G} {\partial \delta_i\partial \delta_j},
\enq
with diffusion constants matrix $\xi$.
The lognormal field prescription amounts to the replacement of the Gaussian density fluctuation field $\delta$ by a Gaussian log-density field $A = \ln(1 + \delta)$. The most trivial benefit is that in contrast to the Gaussian case, the constraint $\delta = \rho/ {\bar \rho} -1 \ge -1$ is ensured in a sensible way. Mass conservation (the condition $\av{\delta} = 0$) is also ensured by forcing a nonzero mean $\av{A} = -\sigma^2_A/2$, where $\sigma^2_A$ is the variance of the log-density field. We set however the mean of the A field to vanish in what follows, for notational convenience. This is irrelevant to our results, the interested reader can introduce the correct value in each step at the cost of extra ink.



\section{N-point correlators in the lognormal field}
We work with an arbitrary number $d$ of pixels with coordinates $x_i$. The two-point function of the Gaussian A field on these pixels is the finite matrix $\xi_{A,ij} = \xi_A(x_i-x_j)$. We do not write explicitly a time dependence, keeping in mind $\xi_A \propto t$.
For lognormal field  statistics, the following convenient, compact equation holds true for a vector $\vecn$ with entries $n_i$ taking arbitrary values \citep{CarronNeyrinck2012} ($\av{A} = 0$),
\beq \label{mn}
\av{\prod_{i = 1}^d \lp 1+ \delta(x_i)\rp^{n_i}}  = \exp \lp\frac 12 \vecn^T \xi_A \cdot \vecn\rp.
\enq
The standard hierarchy of $N$-point moments is described by  the set of such moments for which the entries $n_i$ take all possible non-negative integer values. We extend the hierarchy, allowing negative values as well: the vector $\vecn$ ('multiindex' henceforth) runs in what follows over lattice points $\mathbb{Z}^d$ .The reason will be clear in a moment. The set of observables we consider is thus
\beq \label{hierarchy}
m_\vecn : = \av{\prod_{i = 1}^d \lp 1+ \delta(x_i)\rp^{n_i}} ,\quad \textrm{with  } \vecn \in \mathbb Z^d
\enq 
Let further be for each pair of multiindices $\vecn,\vecm$ the  matrix element $M_{\vecn\vecm}$ be defined through $M_{\vecn\vecm} = m_{\vecn + \vecm}$. The matrix element $M_{\vecn\vecm} - m_\vecn m_\vecm$ is the covariance between correlators $m_\vecn$ and $m_\vecm$. As a meaningful information measure we consider Fisher information  \citep[e.g.]{TegmarkEtal1997}, an information measure associated to arbitrary model parameters $\alpha,\beta$, that quantifies how well these parameters may be inferred from the data under consideration. Our aim is to evaluate the information contained in the hierarchy \eqref{hierarchy} of observables, defined by
\beq \label{Fab}
F^{\textrm{hierarchy}}_{\alpha \beta} = \sum_{\vecn,\vecm \textrm{ in }\mathbb Z^d}\frac{\partial m_\vecn}{\partial \alpha} \lb M^{-1}\rb_{\vecn\vecm}\frac{\partial m_\vecm}{\partial \beta},
\enq
which is at most that contained in the entire (density or log-density) field given by
\beq \label{Ftot}
F^{\textrm{total}}_{\alpha\beta} = \int d^dA\frac{\partial_\alpha p_G(A) \partial_\beta p_G(A)}{p_G(A)}.
\enq
The former reduces to the latter when the probability functional is uniquely defined by the moment hierarchy (see \citep{Carron2011} for a discussion). $F_{\alpha\beta}^{\textrm{total}}$ is in our case the information of a Gaussian random field, given by a well-known formula \citep{VogeleySzalay1996,TegmarkEtal1997} collecting the independent Fourier modes of the field; we will not need here its specific form however.
\subsection{Inverse cosmic covariance matrix}
We now obtain the inverse matrix $M^{-1}$. This can be performed elegantly with this multindex notation as follows. We first note the simple transformation
\beq
\begin{split}
&\frac 12 \lp \vecn + \vecm \rp^T \xi_A \cdot \lp  \vecn + \vecm  \rp =\\& \vecn^T \xi_A\cdot \vecn + \vecm^T \xi_A\cdot \vecm 
  -\frac 12 \lp \vecn - \vecm \rp^T \xi_A \cdot \lp  \vecn - \vecm  \rp.
\end{split}
\enq
We then observe that
\beq \label{Tmatrix}
\exp \lb   -\frac 12 \lp \vecn - \vecm \rp^T \xi_A \cdot \lp  \vecn - \vecm  \rp \rb = \frac{1}{m_{\vecn-\vecm}}.
\enq
Using these two relations in Eq. \eqref{mn}, we can write
\beq \label{fun}
m_{\vecn + \vecm} =\frac{ m^2_\vecn m^2_\vecm }{m_{\vecn-\vecm}}.
\enq
We then define the matrix $\T_{\vecn\vecm} = 1/{m_{\vecn-\vecm}}$. This matrix is invariant under lattice translations $\vecn,\vecm \rightarrow \vecn + \vecr,\vecm + \vecr$, and this allows its inversion below through Fourier transformation. Writing relation \eqref{fun} in terms of the matrices $M$ and $\T$, and inverting on both sides, results in
\beq \label{Minv}
\lb M^{-1} \rb_{\vecn\vecm} = \frac{1}{m_\vecn^2}\frac{1}{m_\vecm^2} \lb \T^{-1} \rb_{\vecn\vecm}.
\enq
It remains to invert  $\T$. To that aim, we introduce its Fourier transform. Let $\theta = (\theta_1,\cdots,\theta_d)$, and
\beq \label{pw}
\tlambda = \frac{1}{(2\pi)^d} \sum_{\vecn \textrm{ in } \mathbb Z^d} \frac{1}{m_\vecn} e^{-i \vecn\cdot \theta}.
\enq
From \eqref{Tmatrix}, this series is absolutely well behaved and we need not worry about convergence issues. Inverting this relation gives the Fourier representation of $\T$,
\beq
\T_{\vecn\vecm} = \frac 1{m_{\vecn-\vecm}} = \int_{\mathbb T^d} d^d\theta\: \tlambda e^{i \theta\cdot (\vecn -\vecm)}.
\enq
The region of integration is for all components $\theta_i$ in $[-\pi , \pi]$ (or any interval of length $2\pi$), a short hand notation for which is the $d$-torus $\mathbb T^d$.
This provides us with the inverse matrix :
\beq \label{Tinv}
\lb \T^{-1} \rb_{\vecn\vecm} =\int_{\mathbb T^d} d^d\theta\: \frac 1 {\tlambda} e^{i \theta\cdot (\vecn -\vecm)}.
\enq
A last intermediate result is required before the evaluation of the information matrix. Consider the impact of a parameter $\alpha$ on $\tlambda$. From its Fourier representation follows directly
\beq \label{suma}
\frac{\partial \tlambda}{\partial \alpha} = -\sum_{\vecn \textrm{ in } \mathbb Z^d} \frac{\partial m_\vecn }{\partial \alpha}\frac 1{m_\vecn^2}e^{i \vecn\cdot \theta} .
\enq
We now reach the main result of this Letter. From Eqs. \eqref{Fab},\eqref{Minv},\eqref{Tinv} and  \eqref{suma} we obtain a remarkably simple expression
\beq
F_{\alpha \beta}^{\textrm{hierarchy}} = 
 \int_{\mathbb T^d} d^d\theta \frac{\partial_\alpha \tlambda\partial_\beta \tlambda}{\tlambda} = F^{\textrm{total }p_W}_{\alpha \beta}.
\enq
It has exactly the form \eqref{Ftot} of the total information content not of the Gaussian field $A$ but of a new field $\theta$, on the torus, obeying the PDF $\t$. 
\subsection{The 'Wrapped' Gaussian field}
We turn to the justification of our our interpretation of $\t$ as a new random field PDF and its physical meaning. We derive another useful representation of $\t$, that makes its connexion to $p_G$ clearer. Written in full, Eq. \eqref{pw} gives the following \beq \label{pw2}
\tlambda = \frac 1 {(2\pi)^d} \sum_{\vecn \textrm{  in  } \mathbb Z^d} \exp \lp -\frac 12 \vecn ^T \xi_A \cdot \vecn - i\theta\cdot \vecn\rp.
\enq
We transform the first exponential factor using the standard Gaussian integral formula
\beq
e^{-\frac 12 \vecn ^T \xi_A \cdot \vecn } = \int_{\mathbb R^d}\frac{d^dA}{(2\pi)^{d/2}\sqrt{\det \xi_A}} \:e^{  -\frac 12 A^T\xi_A^{-1}\cdot A + iA \cdot\vecn  }.
\enq
The sum over the multindices in Eq. \eqref{pw2} reduces now to the $d$-dimensional Dirac-comb
\beq
\frac 1 {(2\pi)^d} \sum_{\vecn \textrm{  in  } \mathbb Z^d}e^{i\vecn \cdot \lp x -\theta \rp }  = \sum_{\vecn \textrm{  in  } \mathbb Z^d } \delta^D\lp2 \pi\vecn  -(x-\theta)\rp.
\enq
The Dirac delta allows us to undo the $x$-integrations, with the result
\beq \label{pw3}
\tlambda = \frac{1}{(2\pi)^{d/2}\sqrt{\det \xi_A}} \sum_{\vecn \textrm{  in  } \mathbb Z^d }e^{  -\frac 12 (\theta - 2\pi \vecn)^T\xi_A^{-1} \cdot(\theta - 2\pi \vecn) }.
\enq
This makes the positivity of $\t$ obvious. On the other hand, representation \eqref{pw2} tells us that $\t$ integrated on the torus gives unity. This justifies fully our interpretation of $\t$ as a random field PDF. We note from this representation \eqref{pw3} that  $p_W$ reduces to the Gaussian field at early times, given by the term $\vecn = 0$. 
\newline
\indent Now, what is the physical interpretation of this new field ? In fact, $\t$ satisfies exactly the same diffusion equation than $p_G$. Remember that $\xi_A$ evolves proportional to $t$. From its representation \eqref{pw2} follows indeed
\beq
\frac{\partial \t}{\partial t} = \frac 12 \sum_{i,j = 1}^d \xi_{A,ij} \frac{\partial^2 \t}{\partial \theta_i \partial \theta_j},
\enq
where $\xi_A$ in this equation is again the two-point function at $t = 1$.
From representation \eqref{pw3}, initial conditions are the same, the only difference is topological: the field $\theta$ diffuses on the torus. It is physically obvious that the topological constraint must change the longterm behavior of the field. In particular we expect $\t$ to reach homogeneity, as soon as, say, $\sigma_A \sim \pi$. From representation \eqref{pw2} holds indeed that for any shape of two-point function holds at late times
\beq
\tlambda \rightarrow \frac{1}{(2\pi)^d}, \textrm{  and  } \frac{\partial\tlambda}{\partial \alpha}\rightarrow 0.
\enq
This is a rigorous demonstration that the joint information of the hierarchy decays to zero. The critical time $\sigma_A \sim \pi$ is larger than a previous estimate $\sigma_A \sim 0.4$ one of us obtained considering one-point moments only \citep{Carron2011}. This is due to the presence of the inverse powers in the hierarchy considered here.
\newline
\indent
\section{Final words}
We are now in a position to answer the title's question in the case of the lognormal field. A rigorous statement is as follows. In the log-density field decomposed at each point into a phase angle $\theta$ in $-\pi$ and $\pi$ and an amplitude $k$ (an integer)
\beq
\label{decomp}
\ln(1 + \delta(x)) = \theta(x) + k(x)\:  2\pi,
\enq
the hierarchy \eqref{hierarchy}, including powers and inverse powers, measures only the phase field $\theta$. This may as well be formulated in terms of a lossy nonlinear transformation, with which we conclude. We consider two hypothetical analysts, facing data in form of the lognormal field on $d$ pixels. The first extracts the hierarchy of correlators \eqref{hierarchy} and discards he original data. The second reconstructs the phase field the log-density in each pixel,
\beq
\ln(1 + \delta(x_i)) \rightarrow \theta(x_i). 
\enq
This can be pictured as wrapping the real line on the unit circle. Following the rules of probability theory, this new data is described precisely by the wrapped Gaussian field $p_W(\theta)$, Eq. \eqref{pw3}. This transformation makes no difference in the initial conditions. Then, under-densities start to permeate in the most over-dense regions, and vice-versa. Eventually, any memory of the initial field is lost, and the data can perfectly be described by white noise. Yet, both analysts have the same information matrix, and are going to reach identical conclusions when performing inference on parameters from their data.
\newline
\indent
\revision{Sourcing this effect is the rapid growth of the moments across the hierarchy, as the density field evolves into the cosmic web. Clearly, the matter density is not exactly lognormal, but, for the reasons exposed in the introduction, it must also be affected by this phenomenon. Trying to understand it for a more realistic model might well be a rather formidable task. This is left for future investigations.}
\newline
\indent
The research leading to these results has received funding from the European
Research Council under the European Union's Seventh Framework Programme
(FP/2007-2013) / ERC Grant Agreement No. [616170].
J.C. thanks Nick Kaiser, Antony Lewis and Peter Coles for comments on the manuscript.
\bibliographystyle{mn2e}
\bibliography{bib}

\begin{thebibliography}{}

\bibitem[\protect\citeauthoryear{Albrecht et~al.,}{Albrecht
  et~al.}{2006}]{Albrecht:2006um}
Albrecht A.,  et~al., 2006, ArXiv Astrophysics e-prints astro-ph/0609591

\bibitem[\protect\citeauthoryear{{Ata}, {Kitaura} \& {M{\"u}ller}}{{Ata}
  et~al.}{2015}]{AtaEtal2015}
{Ata} M.,  {Kitaura} F.-S.,    {M{\"u}ller} V.,  2015, \mnras, 446, 4250

\bibitem[\protect\citeauthoryear{{Bernardeau}}{{Bernardeau}}{1994}]{Bernardeau1994}
{Bernardeau} F.,  1994, \aap, 291, 697

\bibitem[\protect\citeauthoryear{{Bernardeau}, {Colombi}, {Gazta{\~n}aga} \&
  {Scoccimarro}}{{Bernardeau} et~al.}{2002}]{BernardeauEtal2002}
{Bernardeau} F.,  {Colombi} S.,  {Gazta{\~n}aga} E.,    {Scoccimarro} R.,
  2002, \physrep, 367, 1

\bibitem[\protect\citeauthoryear{{Bernardeau} \& {Kofman}}{{Bernardeau} \&
  {Kofman}}{1995}]{BernardeauKofman1995}
{Bernardeau} F.,  {Kofman} L.,  1995, \apj, 443, 479

\bibitem[\protect\citeauthoryear{{Carron}}{{Carron}}{2011}]{Carron2011}
{Carron} J.,  2011, \apj, 738, 86

\bibitem[\protect\citeauthoryear{{Carron} \& {Neyrinck}}{{Carron} \&
  {Neyrinck}}{2012}]{CarronNeyrinck2012}
{Carron} J.,  {Neyrinck} M.~C.,  2012, \apj, 750, 28

\bibitem[\protect\citeauthoryear{{Carron} \& {Szapudi}}{{Carron} \&
  {Szapudi}}{2015}]{CarronSzapudi2015}
{Carron} J.,  {Szapudi} I.,  2015, \mnras, 447, 675

\bibitem[\protect\citeauthoryear{{Carron}, {Wolk} \& {Szapudi}}{{Carron}
  et~al.}{2014}]{CarronEtal2014}
{Carron} J.,  {Wolk} M.,    {Szapudi} I.,  2014, \mnras, 444, 994

\bibitem[\protect\citeauthoryear{{Coles} \& {Jones}}{{Coles} \&
  {Jones}}{1991}]{ColesJones1991}
{Coles} P.,  {Jones} B.,  1991, \mnras, 248, 1

\bibitem[\protect\citeauthoryear{{Fry}}{{Fry}}{1984}]{Fry1984b}
{Fry} J.~N.,  1984, \apj, 279, 499

\bibitem[\protect\citeauthoryear{{Hilbert}, {Hartlap} \& {Schneider}}{{Hilbert}
  et~al.}{2011}]{HilbertEtal2011}
{Hilbert} S.,  {Hartlap} J.,    {Schneider} P.,  2011, \aap, 536, A85

\bibitem[\protect\citeauthoryear{{Kayo}, {Taruya} \& {Suto}}{{Kayo}
  et~al.}{2001}]{KayoEtal2001}
{Kayo} I.,  {Taruya} A.,    {Suto} Y.,  2001, \apj, 561, 22

\bibitem[\protect\citeauthoryear{{Kitaura}, {Jasche} \& {Metcalf}}{{Kitaura}
  et~al.}{2010}]{KitauraEtal10}
{Kitaura} F.-S.,  {Jasche} J.,    {Metcalf} R.~B.,  2010, \mnras, 403, 589

\bibitem[\protect\citeauthoryear{{Peebles}}{{Peebles}}{1980}]{Peebles1980}
{Peebles} P.~J.~E.,  1980, {The large-scale structure of the universe}

\bibitem[\protect\citeauthoryear{{Szapudi}}{{Szapudi}}{2009}]{Szapudi2009}
{Szapudi} I.,  2009, in {Mart{\'{\i}}nez} V.~J.,  {Saar} E.,
  {Mart{\'{\i}}nez-Gonz{\'a}lez} E.,   {Pons-Border{\'{\i}}a} M.-J.,  eds,
  Lecture Notes in Physics, Berlin Springer Verlag Vol. 665, Data Analysis in
  Cosmology. pp 457--492

\bibitem[\protect\citeauthoryear{Takahashi, Soma, Takada \& Kayo}{Takahashi
  et~al.}{2014}]{Takahashi:2014yqa}
Takahashi R.,  Soma S.,  Takada M.,    Kayo I.,  2014, Mon. Not. Roy. Astron.
  Soc., 444, 3473

\bibitem[\protect\citeauthoryear{{Tegmark}, {Taylor} \& {Heavens}}{{Tegmark}
  et~al.}{1997}]{TegmarkEtal1997}
{Tegmark} M.,  {Taylor} A.~N.,    {Heavens} A.~F.,  1997, \apj, 480, 22

\bibitem[\protect\citeauthoryear{{Valageas}}{{Valageas}}{2002}]{ValageasII}
{Valageas} P.,  2002, \aap, 382, 412

\bibitem[\protect\citeauthoryear{{Vogeley} \& {Szalay}}{{Vogeley} \&
  {Szalay}}{1996}]{VogeleySzalay1996}
{Vogeley} M.~S.,  {Szalay} A.~S.,  1996, \apj, 465, 34

\bibitem[\protect\citeauthoryear{{Weinberg}, {Mortonson}, {Eisenstein},
  {Hirata}, {Riess} \& {Rozo}}{{Weinberg} et~al.}{2013}]{WeinbergEtal2013}
{Weinberg} D.~H.,  {Mortonson} M.~J.,  {Eisenstein} D.~J.,  {Hirata} C.,
  {Riess} A.~G.,    {Rozo} E.,  2013, \physrep, 530, 87

\end{thebibliography}
\end{document}